# Low inhomogeneous broadening of excitonic resonance in MAPbBr$_3$ single crystals

O.A. Lozhkina[1*], V.I. Yudin[1], A.A. Murashkina[1], V.V. Shilovskikh[1], V.G. Davydov[1], R. Kevorkyants[1], A.V. Emeline[1], Yu.V. Kapitonov[1], D.W. Bahnemann[1,2]

[1] Saint-Petersburg State University, ul. Ulyanovskaya 1, Saint-Petersburg, 198504, Russia.

[2] Leibniz University of Hannover, Callinstrasse 3, Hannover, 30167, Germany.

[*] Corresponding author.

Corresponding author's contact details:

e-mail address:   st040462@student.spbu.ru

postal address:   Saint-Petersburg State University, ul. Ulyanovskaya 1, Saint-Petersburg, 198504, Russia.

**Abstract**

We present optical study of MAPbBr$_3$ single crystals grown from solution by inverse temperature crystallization method. Photoluminescence temperature dependence of narrow and isolated exciton resonance showed encouragingly low inhomogeneous broadening $\Gamma \approx 0.5$ meV, which is comparable to those of MBE-grown III-V heterostructures, indicating high quality of the solution-grown perovskite. Excitonic origin of the resonance was proved by its superlinear pump intensity dependence in contrast to the linear behavior of the defect-assisted recombination bands. In addition, for the first time phonon replicas were resolved in MAPbBr$_3$ photoluminescence spectra and attributed to the known lattice vibrational modes by comparison with Raman scattering spectra.

**Table of Contents (TOC) Graphic**

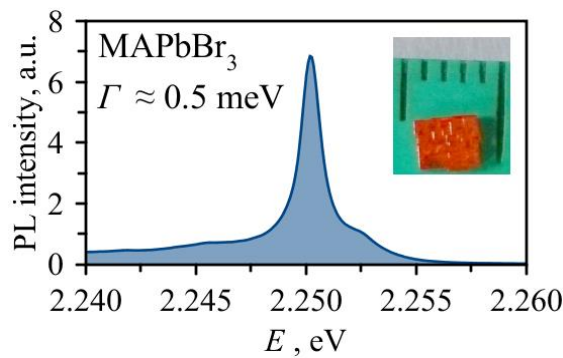

1. **Introduction**

Lead-halide perovskites synthesized by low-temperature solution process received considerable attention as a promising photo-absorber material for low-cost high-efficiency solar cells in recent years. Currently, the perovskite solar cells achieve conversion efficiencies up to 20% [1].

Perovskites demonstrate superior optoelectronic properties such as extremely high luminescence efficiencies, caused by a band-to-band transition due to the material's direct bandgap character [2], and long charge carrier diffusion lengths [3]. Photoluminescence (PL) of perovskites exhibits no Stokes shift between absorption edge and PL that enables efficient photon recycling and significantly improves solar cells efficiencies [4].

At high temperatures lead-halide perovskites $APbX_3$ possess cubic crystal symmetry. Therein, lead cations are surrounded by six halogen anions forming $[PbX_6]$ octahedra. The cations A are located at the positions (1/2; 1/2; 1/2) between the arrays of the three-dimensional network of the octahedra. In contrast to inorganic cations A, the organic ones like methylammonium $CH_3NH_3^+$ (MA) may have different orientations around normal A position. With temperature decrease crystal symmetry lowers to tetragonal, $[PbX_6]$ turn with keeping nearly octahedral shape and the disorder of the organic cation orientation diminishes. Further cooling of the perovskites changes their lattice symmetry to orthorhombic with octahedra tilting and fixes positions of the organic cations [5].

Electronic properties of lead-halide perovskites is a result of interplay between electronic energetic levels of lead and halogens in $[PbX_6]$ octahedra: the main contributions into valence band come from halogen p and Pb 5s lone pair antibonding orbitals, while conduction band consists mostly of halogen p and Pb 5p bonding orbitals [6]. This leads to defect tolerance of perovskite electronic structure. That is, the formation energy of low-lying defects, i.e. Shockley-Read-Hall traps, is high and they do not affect perovskite's optical properties [7]. This, in turn, enables facile synthesis of lead halide perovskites.

The unique luminescence properties of lead-halide perovskites open new perspectives for fundamental optical science and device applications. Recently, a number of optoelectronic devices were proposed which could make use of such crystals. Those include X-ray [8] and γ-detectors [9], optical imaging device [10] etc. A thorough understanding of fundamental optical properties and their relation to the device performance is needed. Single crystals of perovskites are most suitable for fundamental studies on optical and vibrational properties because they allow for elimination of grain boundaries related phenomena. Several recent works were devoted to solution-processed perovskite single crystals of large size (up to a few mm) and high optical quality [11-14]. The knowledge of optical properties of the perovskites gained from such experiments is inevitable for further studies on more complex systems e.g. nanocrystals, powders, and thin films.

In this work we present optical study of high-quality methylammonium lead tribromide MAPbBr$_3$ single crystals.

## 2. Synthesis and characterization

MAPbBr$_3$ single crystals were synthesized following the inverse temperature crystallization method [15] from dimethyl sulfoxide (DMSO) solution. For this purpose, the solution of 1 mol/l of MABr and 1 mol/l of PbBr$_2$ in DMSO was prepared, filtered, and kept in autoclave at 70º C for 4 hours. The precipitated single crystals were isolated and stored in desiccator.

Structural characterization of the solution-grown single crystals was carried out using Scanning Electron Microscope (SEM) Hitachi S-3400N. Their elemental composition was investigated by means of Energy-Dispersive X-ray (EDX) spectrometer Oxford Instruments X-Max 20, while the crystal's structure was confirmed by Electron Backscatter Diffraction (EBSD) using AZtecHKL Channel 5. EDX spectrum in the Figure 1a proves accordance of the sample composition with MAPbBr$_3$ empirical formula. EBSD pattern (Fig. 1b) was fitted by a crystal model based on MAPbBr$_3$ XRD data from paper [16]. Good fit (Fig.1c) proves Pm3m symmetry of the grown MAPbBr$_3$ perovskite crystals.

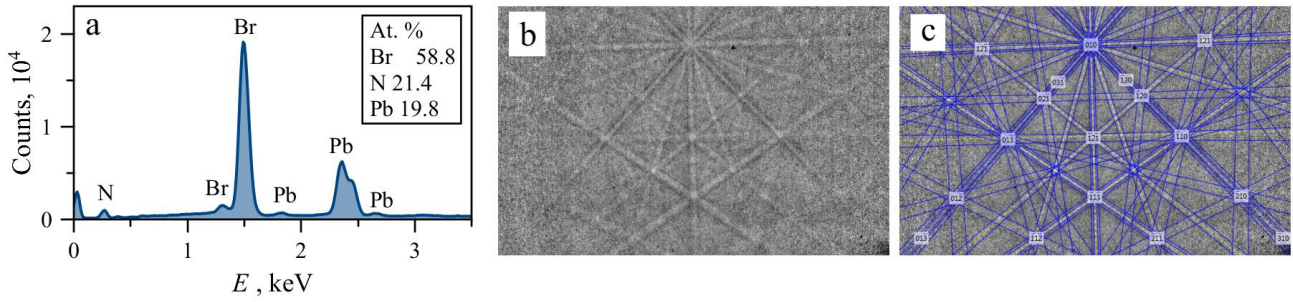

Figure 1.   EDX spectrum of the MAPbBr$_3$ single crystal (a), EBSD pattern of the same crystal (b), and pattern fit with the known Pm3m crystal structure of MAPbBr$_3$ (c).

## 3.   Photoluminescence

For conducting photoluminescence (PL) studies, the grown MAPbBr$_3$ sample was cooled in a closed-loop helium cryostat. PL was excited non-resonantly using either 532 nm or 425 nm cw-lasers. The laser spot diameter on the sample and the beam power were 100 μm and 4 mW, respectively. PL signal was recorded following autocollimation scheme and using Horiba iHR550 spectrometer with liquid-nitrogen cooled CCD detector.

The recorded low temperature PL spectrum of the MAPbBr$_3$ single crystal is shown in the Figure 2a. Therein, the most pronounced peak centered at 2.250 eV we assign to recombination of a free exciton (FE). The two orders of magnitude weaker blue-shifted peak at 2.260 eV could be attributed to the band-to-band transition (radiative recombination, RR) based on the perovskite's band gap and the exciton binding energy determined in paper [14]. The red-shifted, broad, and low-intensity bands correspond to the recombination of excitons localized on the defects (Shockley-Read-Hall recombination centers, SRH). Sharp satellites from low-energy side of FE peak are attributed to phonon replicas and will be discussed in the following.

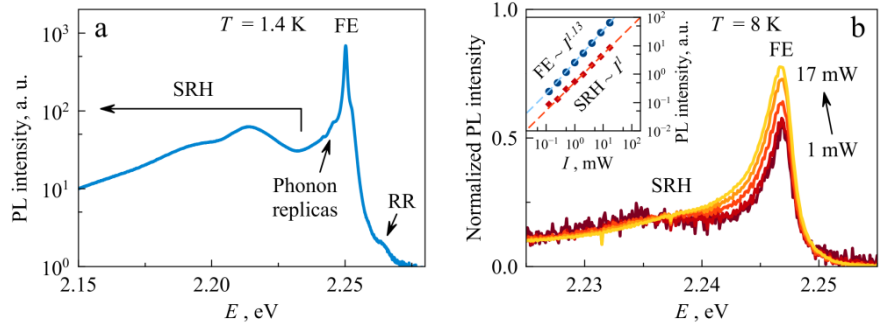

Figure 2.  PL spectrum of MAPbBr$_3$ single crystal (log scale) (a), Normalized pump intensity dependence of PL spectra (b), inset – peaks' intensity dependencies. Assigned peaks: FE – free exciton, RR – radiative recombination, SRH – Shockley-Read-Hall recombination.

Pump intensity dependence of MAPbBr$_3$ PL spectra at T = 8 K is shown in the Figure 2b. FE peak has superlinear dependency (see inset in Fig.2b), which is indicative of its excitonic origin. SRH peak's intensity has linear pump dependency and follows quasi single-particle approximation, because the concentration of charge carriers in this case is considerably lower than that of the charge traps.

Temperature dependence of PL spectra of MAPbBr$_3$ is shown in the Figure 3a. With increasing temperature PL peaks broaden and become blue-shifted due to electron-phonon coupling [14] and thermal expansion of crystal lattice [17]. PL peaks were fitted with Lorentzian and their homogeneous broadening caused by the interaction with phonons was determined. Narrowness of exciton resonance makes it possible to follow homogeneous component down to helium temperatures where this broadening increases linearly with temperature that could be explained by scattering of excitons on acoustic phonons in the Lee-Koteles model [18]. Based on this model we estimate acoustic phonon scattering coefficient in our MAPbBr$_3$ single crystal at 0.07 meV/K. Extrapolation of the phonon-assisted homogeneous broadening to 0 K gives value of inhomogeneous broadening $\Gamma$ equal to 0.5 meV. Such small broadening indicates good optical quality of the solution-grown MAPbBr$_3$ crystal. The obtained $\Gamma$ value is just half an order of magnitude larger than that of the best GaAs epitaxial heterostructures [19].

The Figure 3a demonstrates dependence of PL intensity on a temperature. The measured intensities can be fitted using Arrhenius-like equation [20] taking into account exciton binding energy (Fig. 3b). The latter agrees well with the value found in [14]. Temperature dependence of FE peak and its linear approximation are depicted in the inset of the Figure 3b. The observed rise of energy is unusual for semiconductors, whose band gap energy decreases with temperature-caused interatomic spacing rise [21]. However, the linear dependencies in the inset and their tangents are similar to those of other perovskite materials, e.g. polycrystalline tetragonal and cubic phases of $MAPI_3$, $MAPbI_{3-x}Br_x$ and $MAPbI_{3-x}Cl_x$ [17,22], powders of orthorhombic, tetragonal, and cubic phases of $MAPbBr_3$ and $FAPbBr_3$ [23], and polycrystalline films of $CsSnI_3$ [24].

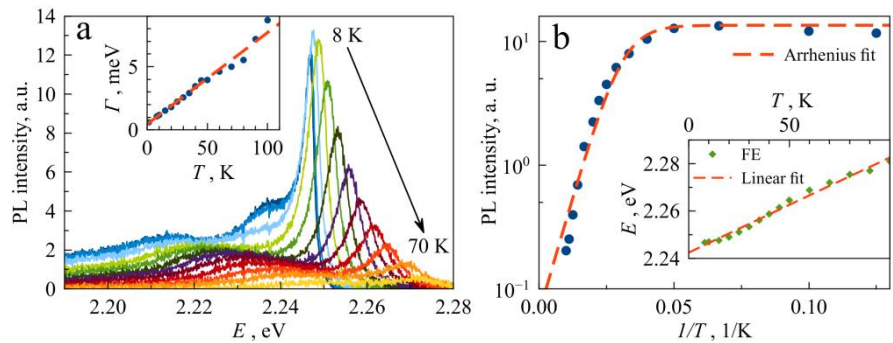

Figure 3. Dependence of PL spectra (a) and PL intensities (b) of the single crystals of $MAPbBr_3$ on a temperature.

### 4. Phonon replicas

In the PL spectrum of the $MAPbBr_3$ single crystals at 1.4 K we observe several weak and narrow bands with the energies lower than that of the FE. We interpret those as phonon replicas, i.e. bands originating from exciton recombination, which is accompanied by partial energy transfer to the $MAPbBr_3$ crystal lattice vibrations (phonons). Since phonon modes in the crystal are quantized the replicas are shifted with respect to the FE resonance by characteristic energy portions.

For the purpose of their identification we have conducted Raman scattering study of the $MAPbBr_3$ single crystal. The Raman scattering spectra were recorded using high resolution spectrometer Horiba T64000 with triple monochromator and excitation laser with a wavelength of 770 nm, i.e. in a transparency range of the sample. In the latter experiment, the sample was cooled down to 7 K using closed-loop helium cryostat. The Figure 4a shows the obtained temperature dependence of Raman scattering spectrum. The most intense peaks are observed at the Raman shifts of 38 cm$^{-1}$, 48 cm$^{-1}$, 67 cm$^{-1}$, and 71 cm$^{-1}$ (at 7 K). Due to the same origin, the peaks at 38 cm$^{-1}$ and 48 cm$^{-1}$ merge at higher temperatures. This is a result of degeneration arising from temperature-driven orthorhombic to higher symmetry tetragonal phase transition. Based on the earlier theoretical analysis of Raman scattering spectra of MAPbX$_3$ [25], the peaks at 38 cm$^{-1}$ and 48 cm$^{-1}$ should correspond to [PbBr$_6$] octahedra twist (TO), at 67 cm$^{-1}$ – to MA pumping around N, and at 71 cm$^{-1}$ – to lurching of MA (LO-like). Comparison of the PL and Raman scattering spectra reveals that the shifts of the weak red-shifted bands off the FE peak (Fig. 4b) correlate with the energies of lattice phonons. Therefore, we conclude that these red-shifted bands are phonon replicas of the free exciton. Observation of these phonon replicas is a result of extremely small inhomogeneous broadening of the excitonic resonance, which, in turn, manifests high optical quality of the studied single crystal.

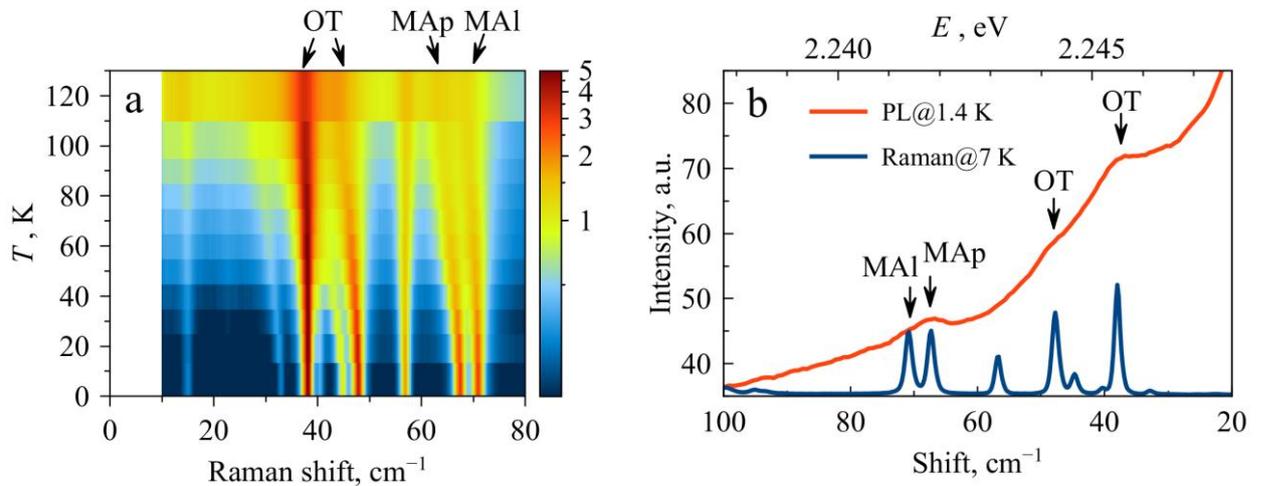

Figure 4. Temperature dependence of Raman scattering spectra of the MAPbBr$_3$ single crystal (a) and correlation between low-temperature Raman and PL spectra revealing phonon replicas (FE peak is set as zero shift) (b). Phonon modes: OT – octahedra twist, MAp – MA pumping around N, MAl – MA lurching.

## 5. Conclusions

We have presented optical study of single crystals of lead-halide perovskite MAPbBr$_3$ grown from supersaturated solution. Extremely low value of inhomogeneous broadening of excitonic resonance in the crystals (~0.5 meV) lets us investigate various low-temperature optical effects. Based on the recorded PL spectra we have succeeded in distinguishing the signals from free excitons and those arising from recombination of weakly localized excitons on defects. We have established the dependence of homogenous broadening on a temperature down to few Kelvins and determined the value of acoustic phonon scattering coefficient (0.07 meV/K). In addition, superior quality of synthesized MAPbBr$_3$ single crystals enabled observation of phonon replicas in PL spectra for the first time. This was confirmed by independent low-temperature Raman scattering experiments.

**Acknowledgment**

The present study was performed within the Project "Establishment of the Laboratory "Photoactive Nanocomposite Materials" No. 14.Z50.31.0016 supported by a Mega-grant of the Government of the Russian Federation. This work was supported by Russian Science Foundation (grant 17-72-10070). The work was carried out using equipment of the resource centers "Nanophotonics", "Geomodel", and "Centre for Optical and Laser Materials Research" of Saint-Petersburg State University.